# The role of vimentin in regulating cell-invasive migration in dense cultures of breast carcinoma cells

Y. Messica[1], A. Laser-Azogui[1], T. Volberg[2], Y. Elisha[2], K. Lysakovskaia[3,4,5], R. Eils[3,4], E. Gladilin[3,4,6,*], B. Geiger[2,*], R. Beck[1,*]

[1] Raymond and Beverly Sackler School of Physics and Astronomy, Tel Aviv University, Tel Aviv, Israel; [2] Department of Molecular Cell Biology, Weizmann Institute of Science, Rehovot, 7610001, Israel; [3] Division of Theoretical Bioinformatics, German Cancer Research Center, 69120 Heidelberg, Germany; [4] BioQuant and IPMB, University of Heidelberg, 69120 Heidelberg, Germany; [5] International Max Planck Research School for Molecular Biology, Georg-August-University Göttingen, 37077 Göttingen, Germany; [6] Leibniz Institute of Plant Genetics and Crop Plant Research, 06466 Seeland, Germany.

**Abstract**

Cell migration and mechanics are tightly regulated by the integrated activities of the various cytoskeletal networks. In cancer cells, cytoskeletal modulations have been implicated in the loss of tissue integrity, and acquisition of an invasive phenotype. In epithelial cancers, for example, increased expression of the cytoskeletal filament protein vimentin correlates with metastatic potential [1,2]. Nonetheless, the exact mechanism whereby vimentin affects cell motility remains poorly understood. In this study, we measured the effects of vimentin expression on the mechano-elastic and migratory properties of the highly invasive breast carcinoma cell line MDA231. We demonstrate here that vimentin stiffens cells and enhances cell migration in dense cultures, but exerts little or no effect on the migration of sparsely plated cells. These results suggest that cell-cell interactions play a key role in regulating cell migration, and coordinating cell movement in dense cultures. Our findings pave the way towards understanding the relationship between cell migration and mechanics, in a biologically relevant context.

Mechanosensing of the pericellular environment plays a vital role in regulating diverse cellular functions, including growth, division, differentiation, morphogenesis, and migration [3–5]. These mechanical cues play key homeostatic roles, and their malfunction is associated with diverse pathological states such as developmental disorders, tumorigenicity and metastasis [6,7]. Researchers have recently taken particular interest in the mechano-elastic properties of cancer cells, showing that cancer cells tend to be softer than benign cells [8]. This same correlation can also be used to distinguish between metastatic and non-metastatic cancer cells, metastatic cells being softer still [9]. The naïve explanation for this correlation is that cell compliance and deformability promote invasion and metastasis by enabling cells to pass through narrow tissue barriers [9].

The cytoskeleton, consisting of microfilaments, microtubules and intermediate filaments (IFs) is the cell's major stress-bearing component, determining its deformability in response to external forces. Microfilament and microtubule networks are primarily composed of F-actin and tubulin, respectively, while the building blocks of IFs comprise versatile groups of proteins that are expressed in a highly tissue-specific and cell type-specific manner [10]. Thus, epithelial cells express diverse cytokeratins, muscle cells express desmin, nerve cells express neurofilaments, glial cells express the glial fibrillary acidic protein, and mesenchymal cells express vimentin [11]. Vimentin is also upregulated in epithelial cells undergoing epithelial-to-mesenchymal transition (EMT) [1,12], alongside downregulation of keratin IFs [13–15]. During EMT, epithelial cells also lose their apical cell–cell junctions and, consequently, their apical-basal polarity, thereby acquiring a migratory capacity, and other mesenchymal characteristics [12]. EMT, which is essential for fundamental developmental processes such as embryogenesis and organogenesis, is also involved in malignant transformation, enabling tumor cells to invade surrounding tissues and eventually form distant metastases [12,16].

The cytokeratin-to-vimentin transition is one of the most prominent and functionally significant processes associated with EMT: this switch underlies the reduction in cell stiffness, an increase in deformability, and augmentation of invasive migration [7,17], all hallmarks of malignant transformation [18]. The direct involvement of vimentin in cell migration and invasiveness is supported by the demonstration that its overexpression in the vimentin-negative, non-invasive MCF-7 breast cancer cell line alters the integrin profile in the cells, and increases their invasiveness [19,20]. Moreover, vimentin expression in epithelial cells induces cell elongation, and loss of cell-cell adhesion [2].

Ample evidence linking vimentin expression to metastasis raises several intriguing questions: What is the mechanism underlying the effect of vimentin on cell migration? Does vimentin directly modulate the migratory machinery of the cells, or is its effect indirect, altering the cell's response to environmental cues?

Using direct physical characterizations, we demonstrate that vimentin expression, combined with cell density, regulates the mechanical properties of MDA231 cancer cells, resulting in their invasive phenotype. The cell density dependence underlying vimentin's effects would suggest that vimentin expression mediates physical and environmental cues, rather than merely affecting the cell motility apparatus. Although some studies show a positive correlation between cell deformability and metastatic potential (migration rate and invasiveness) [9,21], this was not the case with MDA231 cells. We demonstrate that despite being softer than control MDA231 cells, vimentin-deficient cells (MDA231$^{vim-}$) display decreased, rather than elevated, motile and invasive capabilities.

To investigate the possible mechanisms underlying vimentin's regulation of cell mechanics and motility, we performed a series of experiments aimed at evaluating the effects of its expression in various cellular settings, focusing mainly on assays conducted at different cell densities. We studied the effects of vimentin on two pairs of cells two cell lines: MDA-MB-231 and MCF7 cells. MDA-MB-231 (referred to here as MDA231) cells are breast adenocarcinoma epithelial metastatic cells, which naturally co-express both vimentin and keratin IFs (Fig. 1A) [20,22]. MDA231 cells possess a mesenchymal shape, and are commonly used as a model system for a stable EMT state [23]. The second cell line, MCF7, typically expresses keratin IFs only and was used to further examine the effects of vimentin on keratin networks.

To determine the contribution of vimentin expression to cytoskeletal organization within MDA231 cells, we knocked down vimentin expression by means of viral shRNA infection (see Materials and Methods in Supporting Information). The knockdown clones (MDA231$^{vim-}$) display reduced vimentin levels (approximately 0.2% compared to controls; see Supporting Information), with the vimentin being scattered throughout the perinuclear area (Fig. 1B). Notably, the tubulin and actin cytoskeletons remained largely intact in the MDA231$^{vim-}$ cells, similar to those seen in the MDA231 parental cells. Western blot experiments quantitatively confirmed the knockdown of vimentin and, moreover, that the actin network was unaffected (Supporting Information, Section I). In contrast, the cytokeratin cytoskeletal network, which was widely distributed throughout the cytoplasm of MDA231 cells, collapsed into the perinuclear region in MDA231 cells lacking vimentin (MDA231$^{vim-}$), as shown in Fig. 1A.

In the second cell line studied (MCF7), the cytokeratin organization in parental cells forms a typical cage-like network. Following induction of vimentin expression by doxycycline over a 24-hour period (MCF7$^{vim+}$), an extensive vimentin network formed, and the perinuclear keratin network was radically transformed into an extended filamentous structure distributed throughout the cytoplasm (Fig. 1C-D). These results indicate that vimentin expression leads to the dispersal of cytokeratin filaments into the cell periphery. Furthermore, in vimentin's absence (e.g., in epithelial cells, which do not naturally express vimentin, or in cells which underwent EMT and were induced to undergo vimentin knockdown), the cytokeratin network mainly centers around the nucleus.

Vimentin knockdown also reduced the projected area of focal adhesions (FAs), visualized by labeling the cells for several FA-associated proteins (α-actinin, zyxin and vinculin; see Fig. 1E). In addition, we found reduced levels of Snail1 mRNA (Fig. 1F), a transcription factor involved in downregulating E-cadherin and claudins, and upregulating vimentin and fibronectin during EMT [24]. Surprisingly, N-cadherin mRNA levels are increased in vimentin knockdown clones, even though N-cadherin mRNA is associated with EMT and metastatic potential [25–27]. The mRNA levels of other EMT markers; e.g., fibronectin, HMGA2, Zeb1 and Slug, were not significantly altered following vimentin knockdown.

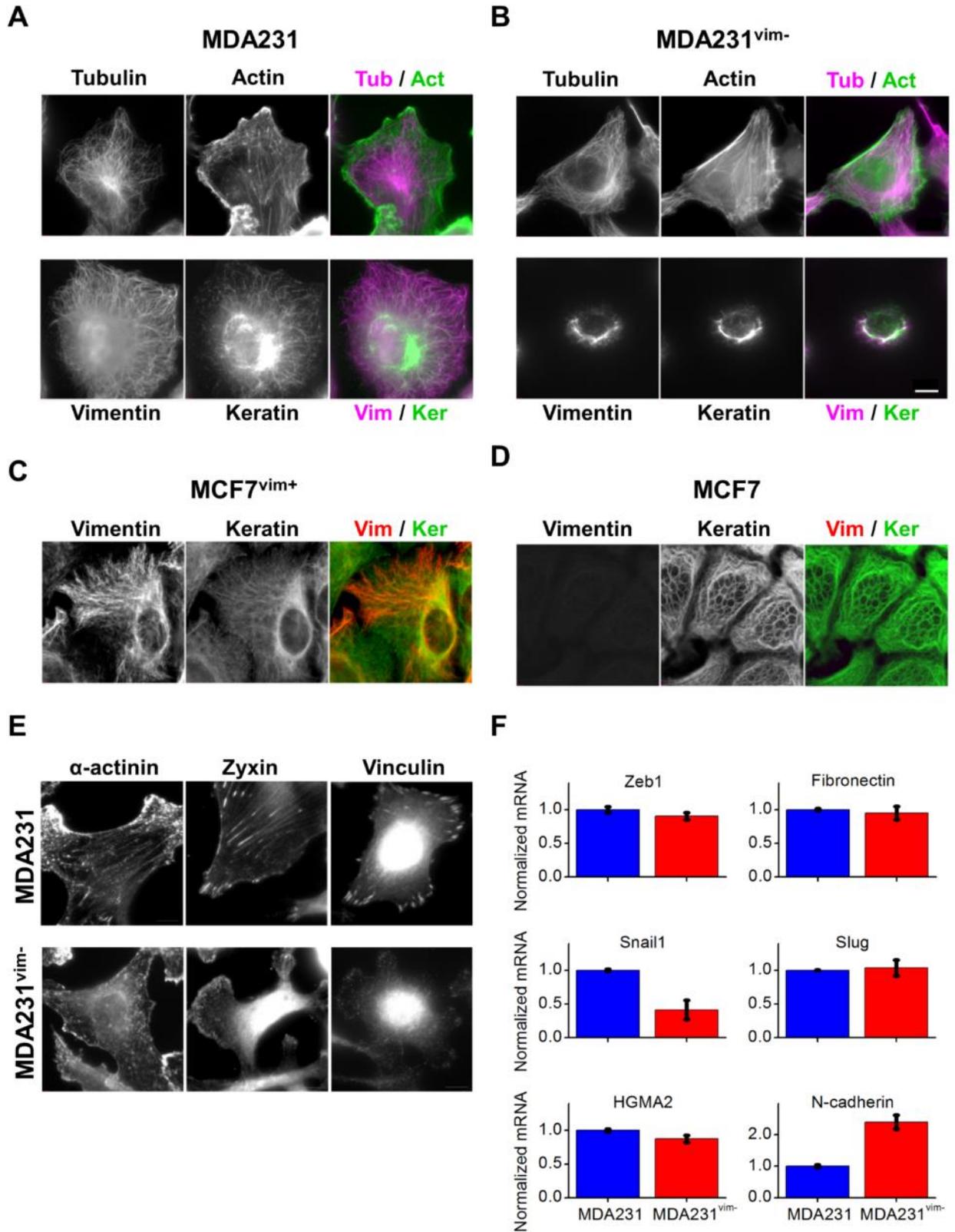

**Figure 1. Immunostaining of cytoskeletal proteins in MDA231 and MCF7 cells, in which vimentin is either expressed, or not expressed.** (A-B) In both MDA231 and MDA231$^{vim-}$ cells, tubulin and actin networks appear similar (upper panels). In MDA231 cells, vimentin and cytokeratin networks span the entire cytoplasm, while in MDA231$^{vim-}$ cells, vimentin expression is diminished, and the keratin network collapses into the perinuclear region. Scale bar: 10 μm. (C-D) Cytokeratin organization in parental MCF7 cells displays a typical cage-like network surrounding the nucleus. Following induction of vimentin expression by doxycycline (namely, MCF7$^{vim+}$ cells), the vimentin filaments form an extensive network, and the keratin network is altered from a cage-like to a dispersed filamentous structure. Addition of doxycycline to parental MCF7 cells had no structural effect on their cytoskeletons. (E) Immunostaining for the adhesion-associated proteins α-actinin, zyxin and vinculin. (F) mRNA levels of EMT markers. Mean values $\pm$ standard error of the mean (SEM) are shown. MDA231: n=3; MDA231$^{vim-}$: n=9.

A more direct effect of vimentin knockdown and keratin network collapse could be observed by directly measuring the cells' rigidity, by means of atomic force microscopy (AFM) and a microfluidics optical stretcher (MOS). Using AFM, the Young's modulus ($E$) of individual cells is extracted by indenting them, and measuring the cell's resistance to the indentation. Comparing the rigidity of MDA231$^{vim-}$ cells to that of control cells indicated that vimentin knockdown induced about 30% reduction in the cells' Young's modulus value, as shown in Fig. 2A. Additional experiments using an MOS confirmed that the vimentin-deficient cells are more compliant (Fig. 2B). In the MOS experiments, the entire cell is stretched, and its overall response to the strain is measured. At the end of the creep phase (denoted by a box), a relative elongation of 2% was detected in MDA231 cells, while in MDA231$^{vim-}$ cells, the relative elongation increased by 25-50%, compared to the MDA231 cells. Both cell lines displayed no significant difference in recovery rate. Similarly, MCF7$^{vim+}$ cells were less stretched by about 15%, compared with parental MCF7 cells (Fig. 2B). These results suggest that the vimentin network, reinforced by the extended keratin network, contributes to cell resistance to deformation.

The next trait we compared in the vimentin knockdown cell lines was cell migration. For this purpose, cells were seeded at low density, and single-cell trajectories were tracked using semi-automated image analysis (see Supporting Information, Materials and Methods, and Supplementary Movie 1). This analysis indicated that there is no statistical difference between the magnitude of migration velocities displayed by the vimentin-containing and the vimentin-lacking cell lines (Fig. 2C), and that cell velocities are exponentially distributed for both cell lines (see Supporting Information, Section II).

Another assay commonly used for assessing cell migration is a "wound-healing" experiment. Briefly, a scratch is mechanically inflicted on a confluent monolayer of cells, and the rate of wound closure is measured. The parameter that is measured here is collective, rather than single-cell migration. In these experiments, the difference between MDA231 and MDA231$^{vim-}$ cells was profound (unlike the migration rates in sparse cultures), with the MDA231 cells closing the wound 50% faster than MDA231$^{vim-}$ cells (Fig. 2D). Notably, we found no difference in proliferation rates between the MDA231 and MDA231$^{vim-}$ cell lines, ruling out the possibility that the difference in wound healing rates could be attributed to differential proliferation (Supporting Information, Section III). Additional migration and invasion assays, which measure the transwell migration of cells plated at high density on filters coated with basement membrane matrix proteins, support this finding; namely, that depletion of vimentin hinders the migration of densely-plated cells (Fig. 2E). Notably, in these assays, the number of cells which successfully migrate through 8 μm pores was counted; yet despite the fact that MDA231$^{vim-}$ cells are softer, and thus should be able to deform more readily, they were slower to migrate and invade the matrix and filter. Our high-density migration assay findings are in agreement with previous reports showing that vimentin enhances densely-plated cell migration rates [18,28].

At this point, it would appear that the migration rate of breast cancer cells in dense cultures (but not sparse cultures) is highly dependent on vimentin expression. Given this sensitivity to culture density, we monitored single-cell motility as a function of cell density by tracking the cells' trajectories in an extended time-lapse microscopy image series. In our analysis, we drew a distinction between two density quantities: *global density*; i.e., the number of cells captured within the field of the view; and *local density* ($\rho_{local}$), which encompasses the cell's "nearest neighbor" cells, within a range of about two cell diameters (see Supporting Information, Section IV for further details).

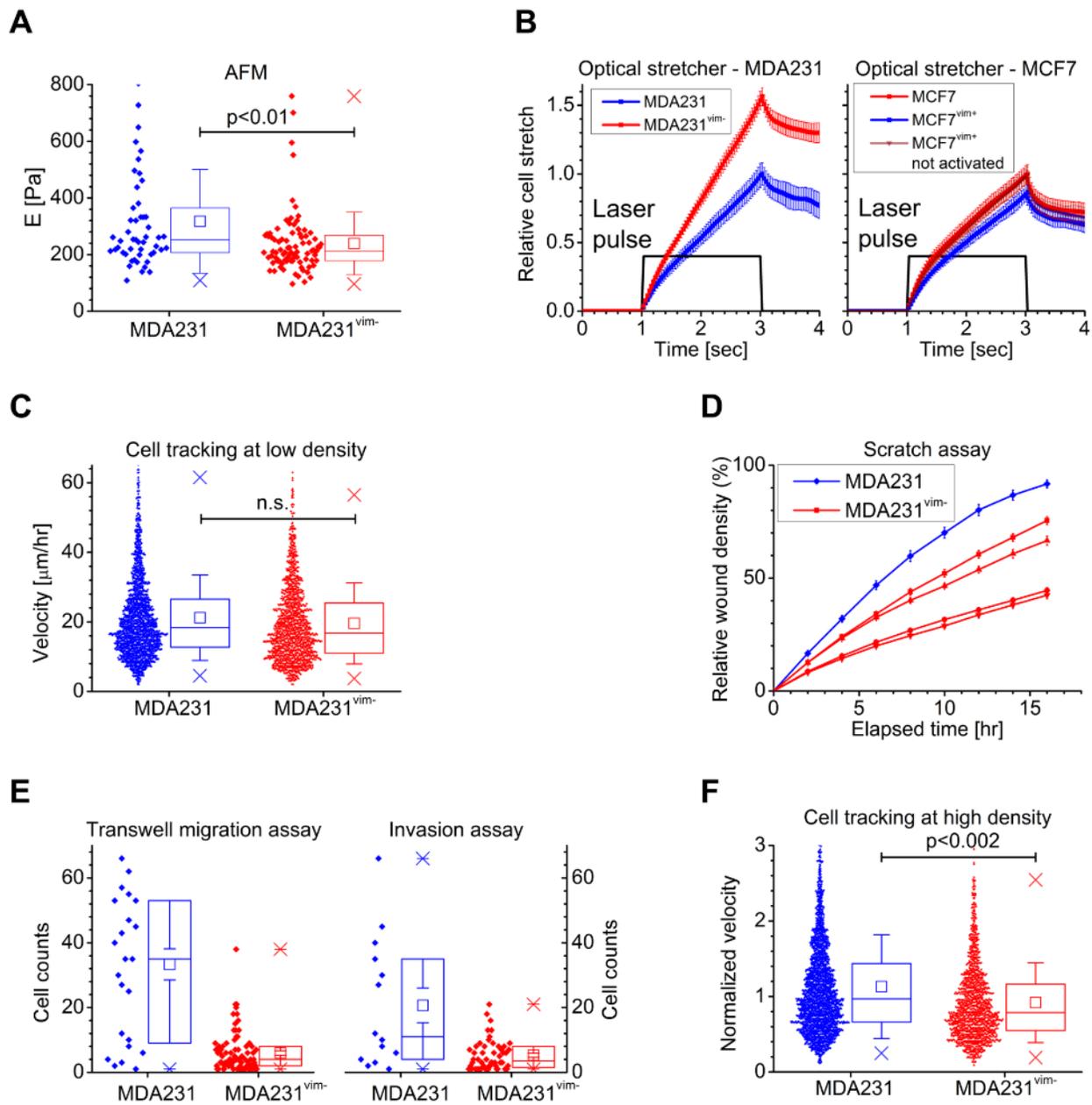

**Figure 2. Elasticity, migration and invasiveness assays**. (A-B) Cell elasticity measurements. (A) Young's modulus distribution, measured by AFM. The Young's modulus in MDA231[vim-] cells is lower than in their vimentin-expressing counterparts. Mean ± SEM values: MDA231: $E = 317 \pm 26\ Pa$ (n=51); MDA231[vim-]: $E = 239 \pm 12\ Pa$ (n=86). (B) Deformability measurements by MOS. MDA231[vim-] cells stretch approximately 50% more than MDA231 cells. Similarly, inducing vimentin expression in MCF7 cells stiffens them by 13%, compared to the MCF7 cells lacking vimentin. Control MCF7[vim+] cells, prior to induction of vimentin by doxycycline, display no such stiffening. Between 300-600 cells were measured per each clone. (C) Velocities of cells seeded at low density (60 $\frac{cells}{mm^2}$) for MDA231 and MDA231[vim-] clones. Median values and standard error calculated using the bootstrap method: MDA231: $v_{LD} =$

$18.43 \pm 0.50 \frac{\mu m}{hr}$ (n=1842); MDA231$^{vim-}$: $v_{LD} = 17.41 \pm 0.52 \frac{\mu m}{hr}$ (n=1368). (D-E) High-density (>400 $\frac{cells}{mm^2}$) bulk migration assays of MDA231 cells and MDA231$^{vim-}$ clones. (D) Wound-healing assay. Each curve represents averaging of 6 different experiments – MDA231 (blue) and 4 clones of MDA231$^{vim-}$ (red). (E) Transwell migration and invasion assays. Each experiment was repeated 3 times, and consisted of imaging 5-10 different fields of view per each clone. These assays, together with the wound–healing assays, show that MDA231 cells are significantly more motile, compared to MDA231$^{vim-}$ cells plated at high density. (F) Velocity distribution at high density, normalized by the velocities at low density. MDA231 cells move slightly more slowly (median value of $\frac{v_{HD}}{v_{LD}} = 0.941 \pm 0.016$, n=2396), while the MDA231$^{vim-}$ cells slow down significantly ($\frac{v_{HD}}{v_{LD}} = 0.772 \pm 0.017$, n=1356).

For each experiment, cells were seeded at low density (~60 $\frac{cells}{mm^2}$), and imaged until the plate reached full confluence. Comparing the ratio between the velocity magnitudes at the end (high density, $v_{HD}$) with those at the beginning of the experiment (low density, $v_{LD}$), we found that MDA231$^{vim-}$ cells slowed by a rate of 23%, while MDA231 cells roughly maintained their velocity (Fig. 2F). In addition, at a relatively high cell density (~400 $\frac{cells}{mm^2}$), the local density strongly correlated with the velocity amplitude for MDA231 cells, as if they were attempting to "escape the crowd". On the other hand, for the MDA231$^{vim-}$ cells, no such correlation was found (Fig. 3A). Supporting evidence comes from evaluating the covariance of the velocity and the local density, $<v\rho_{local}>$, as shown in Fig. 3B. Here, $<>$ represents ensemble and time average for a given experiment. The velocity and local density fields in single frames also exemplify the difference between the cell lines (Fig. 3C, and Supplementary Movie 2). Moreover, in high density plates, we noted that MDA231$^{vim-}$ cells tended to approach their neighbors in closer proximity. However, the difference between the cell lines was found to be rather minor (see Supporting Information, Section V). Notably, even at high density, both cell lines did not form permanent clusters, but rather were found to migrate individually.

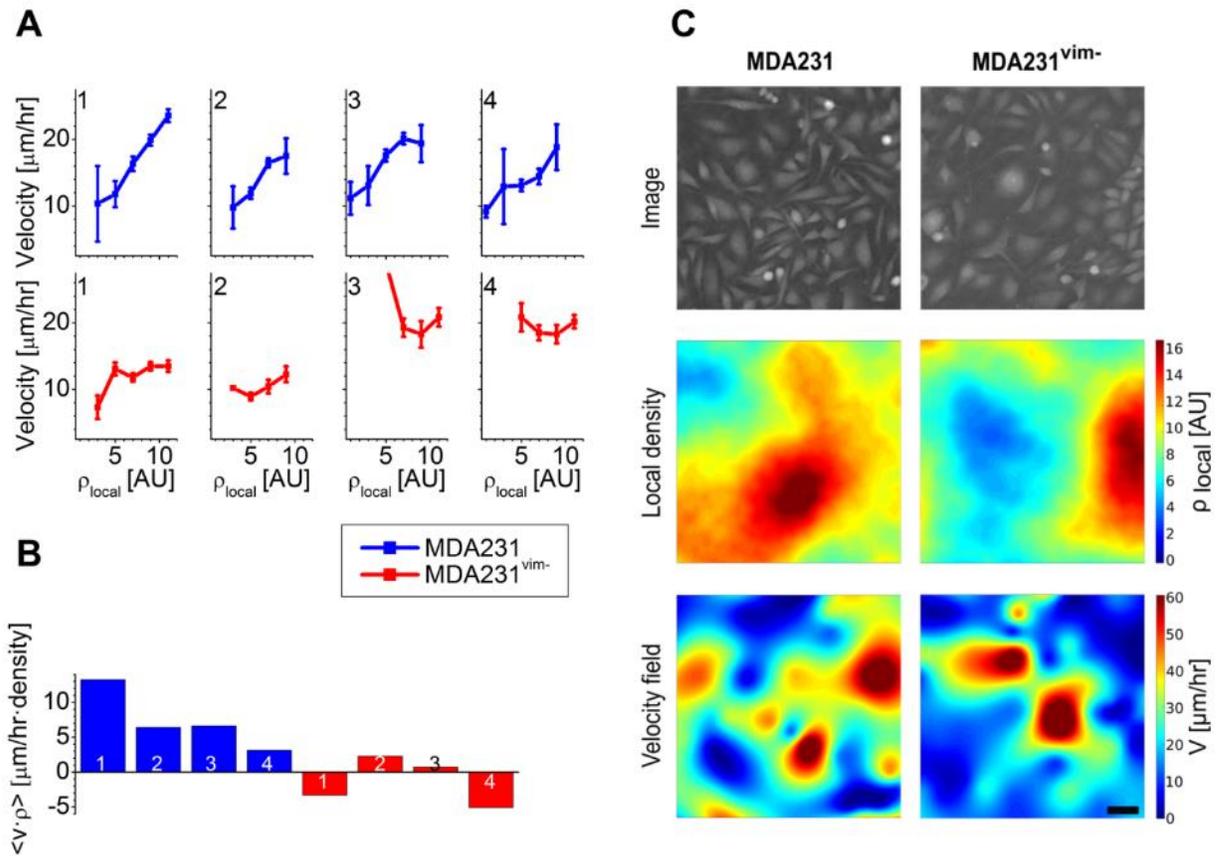

**Figure 3. Single-cell velocity-local density dependence in MDA231 and MDA231 vimentin knockdown (MDA231$^{vim-}$) cells**. (A) Velocity as a function of local density in individual experiments using MDA231 cells at high density (>400 $\frac{cells}{mm^2}$). MDA231 cells show a positive velocity-local density correlation, while MDA231$^{vim-}$ cells display no such correlation. To produce the graph, (velocity, $\rho_{local}$) data points were binned according to $\rho_{local}$ values, and the median velocity in each binned group was taken. (B) Covariance of velocity and local density for the individual experiments shown in (A). MDA231 cells show positive covariance, while MDA231$^{vim-}$ cells do not display a clear trend. (C) Velocity-local density correlation for MDA231 (left) and MDA231$^{vim-}$ cells (right) in a single, high density frame. For each frame (top), the local density field (middle) and the velocity field (bottom) are shown. The velocity fields are interpolated according to the momentary velocity of each cell identified in the frame. Scale bar: 50 μm.

Another property relevant to cell migration is the persistence of the movement. Persistence corresponds to the time duration (or distance) a cell continues to move in a given direction. To extract this quantity, we define the directionality persistence after a time lag $\Delta t$, thus: $C(\Delta t) = \langle \cos(\alpha(t + \Delta t) - \alpha(t)) \rangle$. Here, $\alpha(t)$ is the angle of the cell's velocity vector at time $t$ (see Fig. 4A, inset for illustration), and $\langle \rangle$ stands for averaging over time and the ensemble of the cells. Conceptually, $C(\Delta t)$ decays from unity at $\Delta t = 0$ to zero after a typical time, corresponding to the time it takes for single cells to reorient, and lose their directionality. To obtain a numerical estimate of the persistence time, we fit a double exponential distribution: $C(\Delta t) = Ae^{-\frac{\Delta t}{\tau_1}} + (1 - A)e^{-\frac{\Delta t}{\tau_2}}$, as previously determined to fit experimental data from HaCaT cells [29]. We found that at all global densities, MDA231 cells demonstrate longer persistence time, with $\tau_{2_{MDA231}} > \tau_{2_{MDA231^{vim-}}}$ and $\tau_2$ being the longer characteristic time. Importantly, the difference in persistence time becomes more pronounced at higher global densities (Fig. 4, Table 1). This property can also prove crucial to the advantage of MDA231 over MDA231[vim-] cells in bulk assays, as cells with longer persistence time will display greater net movement, compared to cells that move more erratically.

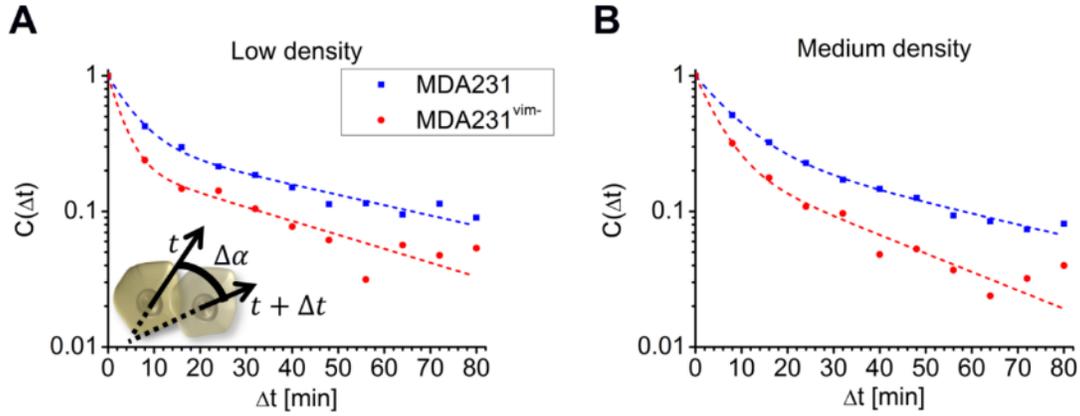

**Figure 4. Persistence time measurements.** Directionality over lag time for MDA231 and MDA231 vimentin knockdown (MDA231$^{vim-}$) cells plated at (A) low (<120 $\frac{cells}{mm^2}$) and (B) medium (200-300 $\frac{cells}{mm^2}$) densities. At low density, data points comprise averages of 6 experiments for MDA231 cells, and 5 experiments for MDA231$^{vim-}$ cells. At medium density, data points comprise averages of 3 experiments for each cell line. Dashed lines: Double exponential fits for the averaged curves. Inset: Schematic of calculation for directionality analysis, derived from measured velocity directions at various time points, separated by intervals of $\Delta t$.

**Table 1.** Fit parameters for MDA231 and MDA231$^{vim-}$ angle autocorrelation data displaying a double exponential decay: $C(\Delta t) = Ae^{-\frac{\Delta t}{\tau_1}} + (1-A)e^{-\frac{\Delta t}{\tau_2}}$

|  | $\tau_1$ [min] | $\tau_2$ [min] | A |
| --- | --- | --- | --- |
| **MDA231 – sparse** | 5.5 | 57.3 | 0.68 |
| **MDA231 – medium** | 8.2 | 54.1 | 0.71 |
| **MDA231$^{vim-}$ - sparse** | 3.1 | 42.3 | 0.78 |
| **MDA231$^{vim-}$ - medium** | 4.7 | 32.0 | 0.77 |

MDA231 cells are more persistent in their movement over time than MDA231$^{vim-}$ cells, as demonstrated by the longer decay times ($\tau_1$, $\tau_2$). The difference between the two cell lines is more pronounced at higher densities.

The results presented above clearly imply that vimentin knockdown in MDA231 cells significantly impairs the cells' migratory activity. To further test the involvement of vimentin in the regulation of cell invasiveness, we conducted an ex vivo infiltration assay, in which two monolayers, one of MDA231 cells (vimentin-expressing or -knockdown) and the other of stromal fibroblasts (Fig. 5A and Supporting Information, Supplementary Movie 3), were seeded separately in two parallel compartments. The monolayers were allowed to migrate and expand, eventually reaching each other (closing the gap), by collective migration. Measurement of migration and invasion speeds indicated that prior to their encounter with the fibroblasts, the MDA231 cells migrated towards the stromal cells ~1.5-2 times faster than their MDA231$^{vim-}$ counterparts (Fig. 5B and Fig. S6), in line with the wound-healing results. After reaching contact, both cell lines slowed their rate of movement; yet the migration of MDA231$^{vim-}$ cells nearly stopped at the boundary between the two cell populations, while the MDA231 cells efficiently infiltrated the stromal monolayer.

Results from our infiltration assays indicate that the motility-density dependence of MDA231 and MDA231$^{vim-}$ cells is not specific to environments involving cells of the same type; rather, that the motility of MDA231$^{vim-}$ cells is impaired in any dense cellular environment.

The experimental results presented in this study show that variations in vimentin levels can alter both the rigidity and the motile behavior of breast cancer cells. However, in contrast with previous reports [30], our findings demonstrate that while the expression of vimentin does not significantly alter the migration of sparsely-plated MDA231 breast cancer cells (Fig. 2C), it *significantly enhances* the collective migration of densely-plated cells (Fig. 2 D-F), suggesting that cell-cell interactions play key roles in regulating migration [31,32].

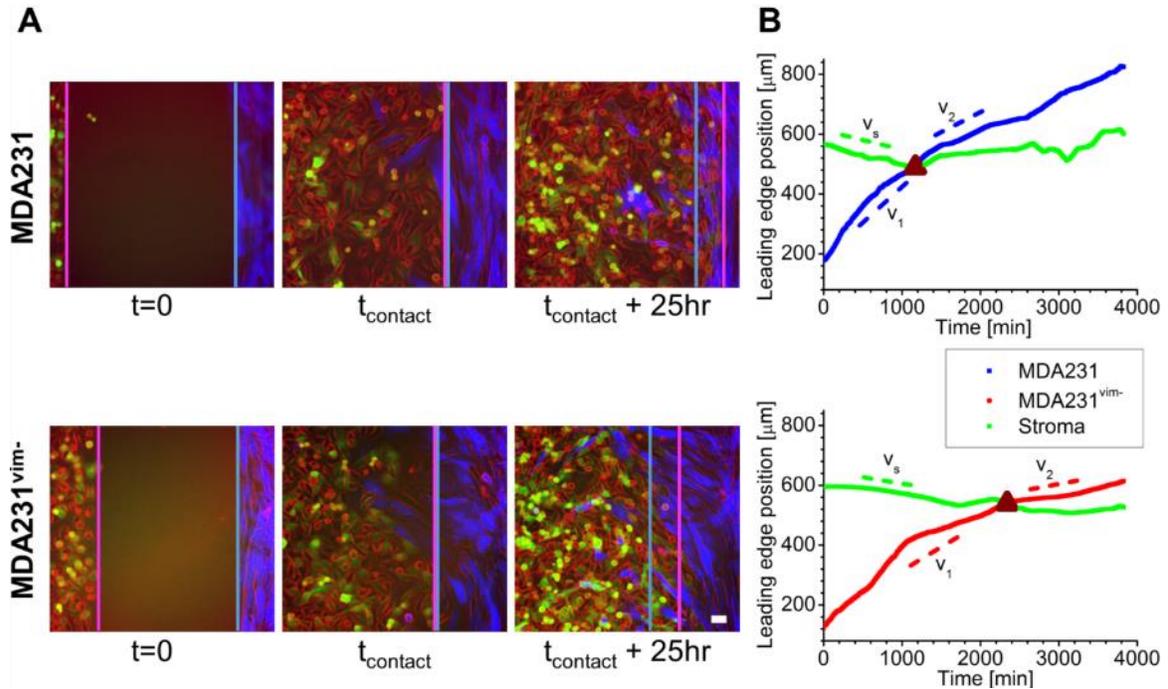

**Figure 5. Infiltration of MDA231 and MDA231$^{vim-}$ cells into stromal fibroblast monolayers**. (A) Snapshots of two fields of MDA231 (green in upper panels) and MDA231$^{vim-}$ (green in lower panels) cells, taken at the initial stage of migration, the time of contact with the stromal monolayer (denoted in blue), and 25 hours post-contact. Magenta and cyan lines mark the monolayer edges as automatically identified (see Materials and Methods). After establishing contact with the fibroblasts, the MDA231 cells infiltrated the stromal cells, while the MDA231$^{vim-}$ cells progressed at a considerably slower rate. Scale bar: 50 μm. (B) Leading-edge progression over time of the MDA231/MDA231$^{vim-}$ (blue/red curves, respectively) and stromal monolayers (green curve). Brown triangles indicate the contact point. Dashed lines are linear fits to the pre-contact step (t=0 up to contact with slope $v_1$) and post-contact progression (from contact up to 8 hours later, with slope $v_2$). The MDA231 cells migrate faster prior to reaching the stromal monolayer, in comparison to the MDA231$^{vim-}$ cells, with a marked difference in rates of persistence seen post-contact. Leading-edge velocities (mean ± SEM; n=8 for both cell lines): $v_1^{MDA231} = 16.0 \pm 1.0 \frac{\mu m}{hr}$, $v_1^{MDA231^{vim-}} = 10.8 \pm 0.6 \frac{\mu m}{hr}$, $v_2^{MDA231} = 7.2 \pm 0.6 \frac{\mu m}{hr}$, $v_2^{MDA231^{vim-}} = 4.2 \pm 1.2 \frac{\mu m}{hr}$ (see also Supporting Information, Fig. S6 for leading-edge velocity distribution).

The direct mechanisms underlying this effect of vimentin on cell migration remain unclear; yet several recent studies presented results that are in line with the data reported here, and proposed possible molecular mechanisms underlying the vimentin-dependent enhancement of cell migration. For example, Chu et al. suggested that co-expression of vimentin and keratin induces increased cytoskeletal interactions at focal adhesions with the extracellular matrix (ECM) which, eventually, affect cell motility [33]. Vuoriluoto et al. proposed that vimentin expression during EMT upregulates the receptor tyrosine kinase Axl which, in turn, contributes to extravasation of breast cancer cells in mice[34].

Reduced intercellular adhesion in vimentin-expressing cells may also result from an indirect effect of vimentin expression on keratin reorganization (Fig. 1A), and consequently on desmosome formation [35]. Danuser's group further demonstrated that vimentin stabilizes the microtubule network, thereby enhancing persistent migration [36], a phenomenon that might be related to the enhanced persistence time we observed in vimentin-expressing cells (Fig. 4).

Another proposed mechanism of vimentin-dependent regulation of cell motility might involve its modulation of the Rac1/RhoA pathway [37], which plays a key role in switching between mesenchymal and amoeboid types of cell migration [38]. Recently, vimentin has been shown to enhance Jagged-mediated Notch signaling [39], which promotes tumor invasion and metastasis [40]. Obviously, some or all of these mechanisms might function in tandem. These previously published studies are in line with the results presented here; yet they do not explain the unexpected dependence of migration rates on *local culture density*, as presented for vimentin-expressing and knockdown cells.

Given the huge complexity of the processes regulating cell migration and invasion, we wish to propose an alternative, purely physical mechanism underlying vimentin-dependent motility-cell density relationships. Since vimentin knockdown softens MDA231 cells (Fig. 2A-B), MDA231$^{vim-}$ cells are, by definition, more susceptible to deformation by external forces. Therefore, each migrating cell from an MDA231$^{vim-}$ culture can easily deform its neighboring cells, rather than reorienting to conform to a vacant intercellular space (see Fig. 6 for a schematic visualization). This characteristic can impair MDA231$^{vim-}$ cell migration under conditions of high cell density, as we indeed observed experimentally. Furthermore, this susceptibility to deformation can account for MDA231$^{vim-}$ cells' loss of polarity, and thus direction of movement, faster than MDA231 cells (Fig. 4). As cells constitute active systems, interactions between cells can give rise to non-trivial dynamics, as also discussed in the context of density-dependent motility [32]. Ultimately, differences in migration speeds of vimentin-containing and vimentin-lacking cells would not be detected at low densities, where cells rarely interact with their neighbors.

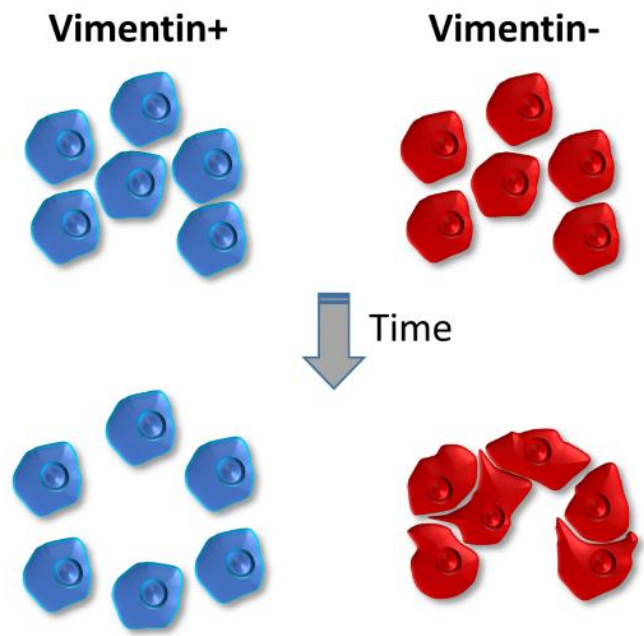

**Figure 6. A scheme of the suggested mechanism underlying vimentin-regulated motility.** MDA231$^{vim-}$ cells are more compliant and readily deformed, compared to MDA231 cells; thus, a migrating MDA231$^{vim-}$ cell can deform surrounding cells, rather than reorienting itself and migrate towards a vacant space.

Moreover, in this study we ruled out the possibility that a coherent mode of migration is a distinguishing trait between the cell lines (Supporting Information, Section VII). Thus, motility of MDA231 cells is enhanced by their neighbors in a manner unrelated to coherent movement (i.e., pulling each other in a common direction). These results indicate that neighboring cells enable each other to develop higher velocities, regardless of their locomotion direction. It should be noted that the average velocity of MDA231 cells does not increase with elevated plate density (Fig. 2F); therefore, additional mechanisms must exist to keep the average velocity at similar values.

Previous studies have shown that the rigidity of a cell's environment affects various cellular properties, motility among them [41–43]. However, the reference to "environment" usually focuses on the ECM. The results presented here show that mechanical cues from neighboring cells can play key roles in environmental signaling. Consequently, the mechanical properties of cells play a dual role as both sensors and inducers of mechanical stress. A body of evidence indicates that increasing substrate stiffness slows the rate of cell migration [3,41,44], rendering the analogy between the mechanical properties of neighboring cells and those of the underlying substrate non-trivial. Tumor growth can produce compressive mechanical stress [45,46] that has been shown to elevate the rates of motility of malignant cell lines, including MDA231 [47]. Density-enhanced motility for the PC-3 prostate cancer cell line was also previously reported [48]. Interestingly, the PC-3 cell line also expresses vimentin, similar to the MDA231 system we studied here. Further research is still needed to evaluate whether the phenomenon of density-velocity correlation in cells is, indeed, widespread, and whether it could originate from sources other than vimentin expression.

In summary, our investigations into the effects of vimentin on cell mechanics and, consequently, the cell's migratory and metastatic potential serve to convincingly illustrate the influence of cell environment on cellular behavior. Our results suggest that the rigidity of the surrounding cells affects the motility of the single cell. Future research should illuminate the mechanical dynamics between cells as they migrate individually; namely, by probing inter-cellular forces and the resulting deformations. There are many interesting biological principles that may be gleaned from such dynamics. Here, these principles give rise to cell density-dependent velocities, as previously reported in other contexts [31,32]. Significantly, vimentin expression regulates this unconventional, density-dependent motility phenotype. A quantitative approach, such as that demonstrated here, should shed light on the mechanisms governing cell motility in different environments.

**Supporting Information**

Materials and Methods, Supporting Sections I-VII, Movies 1-3

**Acknowledgements**

This work was supported by the DKFZ-MOST Program, the Israel Science Foundation (Grant no 550/15), and the Abramson Center for Medical Physics, Tel Aviv University. BG holds the Erwin Neter Professorial Chair in Cell and Tumor Biology. We thank Y. Roichman, Y. Shokef and N. Gov for fruitful discussions. We are also grateful to D. Sprinzak and D. Kaganovich for their kind assistance with the microscopy experiments.

**Corresponding Authors**

Evgeny Gladilin: e.gladilin@dkfz.de

Benny Geiger: benny.geiger@weizmann.ac.il

Roy Beck: roy@tauex.tau.ac.il

# Supporting Information

# The role of vimentin in regulating cell invasive migration in dense cultures of breast carcinoma cells

*Y. Messica, A. Laser-Azogui, T. Volberg, Y. Elisha, K. Lysakovskaia, R. Eils, E. Gladilin, B. Geiger, R. Beck*

**Materials and Methods**

**Cell tracking using a Holomonitor M4 microscope.** For tracking single cell migration, a Holomonitor M4 device [Phase Holographic Imaging PHI AB (publ), Sweden] was used. Cells were seeded at the desired density, and images were taken automatically at intervals of 4 minutes. The accompanying software, HStudio M4, was used to extract coordinates of individual cells, further analysed by software written in-house. To minimize errors due to noise in image acquisition, velocities were calculated from displacements between points at 8-minute intervals, hereby from down-sampled data. Similarly, the angle autocorrelation, local density and flow field were also calculated from the down-sampled data.

**Transwell migration and invasion assays.** Cell transwell migration and invasion kits, CytoSelect™ Cell Migration Assay and CytoSelect™ 24-Well Cell Invasion Assay with basement membrane (Cell Biolabs, Inc., San Diego, CA, USA) were used. In both assays, the plate on which cells were seeded contains polycarbonate membrane inserts with 8 μm pores. Migratory cells are able to pass through the pores of the membrane. In the invasion assay, the upper surface of the membrane was coated with a uniform layer of basement membrane matrix proteins. This basement membrane layer serves as a barrier to discriminate between invasive and non-invasive cells. Invasive cells are able to degrade the matrix proteins, and ultimately pass through the pores of the membrane. In both assays, in the upper well (i.e., insert) we added 105 cells with low-serum (1%) growth medium; fresh complete cell growth medium was added to the bottom well. After 48 hours, cells that migrated through the membrane were counted, following FDA staining.

**Atomic Force Microscopy (AFM).** Cells were cultured in 35-mm culture plates overnight. A Petri Dish Heater (JPK Instruments, Berlin, Germany) was used to maintain cultures at 37ºC, and a flow of 5% $CO_2$/air over the device was used to maintain proper gas exchange. Measurements were performed using [NanoWizardIII](NanoWizardIII) AFM (JPK Instruments) mounted on an Olympus IX81inverted microscope.

Soft AFM probes (spring constant K=0.01-0.08 N/m) with colloidal tips (silicon-dioxide) were used for indentation. Force-distance curves were obtained from several different points on each cell. The local elasticity (Young's modulus) at each point was calculated by fitting the data to Sneddon's solution for a spherical indenter [1]. The Young's moduli for all points on each cell were averaged, to yield the Young's modulus of the cell.

**Immunofluorescent staining.** Cells were cultured on FN-coated cover slips overnight. Where indicated, expression of vimentin in MCF7$^{vim+}$ cells was induced with doxycycline (dox), 5µg/ml for 24 hr. Cells were fixed in either methanol (-20°C) for 2 min, or paraformaldehyde-glutaraldehyde at room temperature for 5 min. Following fixation, cells were permeabilized with 0.5% Triton X-100 for 10 min. Cells were then processed for immunofluorescence with the following antibodies: Anti-Cytokeratin C-11 conjugated to Alexa Fluor 488, 1:50; Anti-Vimentin EPR3776 conjugated to Alexa Fluor 594, 1:100; Anti-tubulin, 1:200; Anti-actinin, 1:100; Anti-vinculin, 1:100; Anti-catenin, 1:100; Anti-zyxin, 1:100. Actin staining was performed with FITC-conjugated palloidin, 1:100. Following 75 min incubation with primary antibodies (/conjugated primary antibodies/ conjugated phalloidin), cells were incubated with secondary antibodies for 1 hr if the primary antibody was unconjugated. Samples were then washed, mounted on microscope slides, and subjected to fluorescence microscopy.

**Wound-healing assays.** Assays were performed on an Incucyte zoom system (Essen BioScience, Ann Arbor, MI, USA) in 96-well ImageLock Plates (Essen BioScience) inside a $CO_2$ incubator. A 96-pin wound-making tool (Essen BioScience) was used to make uniform scratch wounds in all wells. 40,000 cells per well were cultured overnight. Following wound-making, images were automatically acquired every 2 hrs. A Cell Migration Analysis software module (Essen BioScience) was used for analysis.

**Infiltration assays.** Tissue culture dishes were pre-coated with 10 μg/ml FN. MDA-MB-231-GFP WT/vim$^-$, and CCD1069SK-RFP fibroblast cells ($2 \cdot 10^4$) were grown in the two wells of a silicon insert (Ibidi®, Martinsried, Germany), until reaching confluence. The insert was then removed, and the area between the wells was subjected to live-cell imaging, with 15 min intervals between frames, for a total of 64 hrs, using a 10X0.3NA objective.

After obtaining the time-lapse images, the leading edges of the MDA231/MDA231$^{vim-}$ and the stroma monolayers were detected. Each channel was converted to a binary image by thresholding. Image intensity was binned to slices in the horizontal axis. The leading edge of the MDA231/MDA231$^{vim-}$ monolayers was defined as the rightmost slice with an intensity greater than 10% of the maximum intensity, and the leading edge of the stroma was defined as the leftmost slice with an intensity greater than 50% of the maximum intensity.

**Generation of stable vimentin knockdown MDA231 cell clones.** We knocked down vimentin in MDA231 cells (MDA231$^{vim-}$), using an shRNA technique. The resulting cells expressed significantly reduced levels of vimentin. To generate MDA231$^{vim-}$ cells, we constructed lentiviral vectors containing an shRNA sequence, designed using Invitrogen's RNAi designer tool: 5′-GGAAGAGAACTTTGCCGTTGA-3'; the loop sequence between the sense and antisense sequences: 5′-TTCAAGAGA-3'.

The shRNA sequences were cloned under control of a U6 in a pLL3.7 backbone. Viral particles containing either pLL3.7-shVIM or pLL3.7-puro (control) were prepared, and parental MDA231 cells were infected with one of them. Following selection with puromycin, single-cell clones were isolated, and reduced vimentin expression was verified (Figs. 1B, S1).

**Generation of induced vimentin expression MCF7 cell clones.** We successfully generated MCF7 inducible vimentin cells (MCF7$^{vim+}$), which express vimentin upon induction with doxycycline. To generate MCF7$^{vim+}$ cells, we cloned vimentin cDNA into a FuW-TetO lentiviral vector, under control of the tetracycline operator and a minimal CMV promoter. Viral particles containing FUW-TetO-hVimentin and FUW-M2rtTA (a constitutive FUW lentivirus carrying the tetracycline controllable transactivator) were prepared, and parental MCF7 cells were co-infected with the two viruses. Following selection with zeocin, single-cell clones were isolated, and vimentin expression in the presence of doxycycline was verified.

**XTT cell proliferation assays.** An Xtt-based cell proliferation kit was purchased from Biological Industries (Kibbutz Beit-Haemek, Israel). Assays were performed according to the manufacturer's protocol. MDA231 and MDA231$^{vim-}$ cells were seeded at $2.5 \cdot 10^3$, $5 \cdot 10^3$ and $10^4$ cells per well in 96-well plates, with 6 repeats for each cell type/concentration. Cells were allowed to proliferate for 24 hrs; a reaction solution was then added for 3 hours. Dye absorbance intensity was then measured using a plate reader at 500 nm.

**Western blot analysis.** Cells were lysed in modified RIPA buffer (50 mM Tris pH 7.4, 150 mM NaCl, 1 mM EDTA (AppliChem, Darmstadt, Germany), 10 mM NaF, 1% (v/v) NP40, 0.1% sodium deoxycholate, 2 pg/ml aprotinin and 200 pg/ml AEBSF). Cell extracts were cleared by centrifugation at 18,000 g for 10 min at 4°C, and the protein concentration of each sample was measured by the BCA protein assay (Thermo Fisher Scientific, Waltham, MA USA). Equal amounts of lysates were subjected to 10% SDS – PAGE, and subsequently transferred to nitrocellulose membrane (Millipore, Billerica, MA USA). Blots were blocked with blocking buffer for infrared immunoblotting (LI-COR #927-40000) for 1 hr, and co-incubated with primary antibodies against vimentin (Cell Signaling #5741) and actin (Sigma Aldrich #A5441) overnight at 4°C. Secondary antibodies coupled to IRDye infrared dyes (LI-COR #926-32211 and #926-68070) were used for detection with an infrared Odyssey imager (LI-COR). Signal quantification was performed using an ImageQuant system (GE Healthcare, Little Chalfont, UK).

**qPCR for detection of mRNA levels of EMT proteins.** mRNAs were isolated from the cells using an RNeasy Plus Kit (Qiagen, Hilden, Germany) and reverse transcribed to cDNAs with random primers and MultiScribe Reverse Transcriptase (Thermo Fisher, Waltham, MA USA), in accordance with the manufacturer's protocols. cDNAs were used for gene expression analysis in a qPCR reaction with Light Cycler 480 Probes Master (Roche Applied Science, Basel, Switzerland) for 50 cycles. For detection of specific products, the Universal Probe Library (UPL) platform (Roche Applied Science) was used. Glyceraldehyde-3-phosphate dehydrogenase (GAPDH), Glucuronidase Beta (GUSB), and Glucose-6-phosphate dehydrogenase (G6PD) were used as housekeeping genes. Gene expression levels were calculated using the $2^{-\Delta\Delta Ct}$ method.

**High-throughput cell deformability measurements using a microfluidic optical stretcher.** Measurements of cellular deformability were performed using the microfluidic optical stretcher (MOS) system [2] provided by RS Zelltechnik GmbH (Leipzig, Germany). In brief, the MOS generates stretching forces on the boundaries between optically different media, such as the medium and the cell membrane, by exposing suspended cells to two opposing rays of infrared laser light (wavelength=1060nm; power=800mW). The successive deformation of the cell is captured in a time series of 2D phase contrast images. Every cell is monitored 1 sec before and after the 2 sec rectangular laser pulse, resulting in a sequence of totally 120 images. Subsequently, the outer cell contour is segmented, and the diameter of the largest cell axis $d(t)$ is determined for every time step. On the basis of the smoothed time series $d_s(t)$, the relative stretch of the cell is computed as $\varepsilon(t) = \frac{d_s(t)}{d_u} - 1$, where $d_u$ denotes the diameter of the largest axis of unloaded cells. Comparative analysis of population measurements is performed using the parameter-free bootstrap approach [3], which relies on calculation of sample means and corresponding confidence intervals of empirical cell deformability distributions. In our previously published study [4], drug-treated actin- and vimentin-deficient NK cells displayed, on average, 41% and 20% higher optical deformability, in comparison to untreated control populations.

**Statistical analyses.** For the velocity and MOS data, confidence intervals were calculated using the bootstrap method [3]. As the cell tracking experiments showed great variability in cell velocities, median bootstrapped distributions were simulated by pulling an equal amount of data points from each experiment. Statistical significance for AFM data was established, using the Mann-Whitney U-test. P-values for Western blots and proliferation assay data were calculated using the Student's t-test. Box charts (e.g., Fig. 2A) display the 25th-75th percentiles in the box, the median by the dividing line, the mean by the small inner box, the standard deviation by the whiskers, and the 1st and 99th percentiles by the letter X.

## I. Western blots for vimentin knockdown

To quantify the effects of vimentin knockdown on protein expression by the shRNA vector, Western blots for vimentin and actin were performed (Fig. S1A). The knockdown of vimentin was successful, decreasing its expression by 2 orders of magnitude (Fig. S1B). Actin expression levels did not change significantly (p=0.41, Fig. S1C).

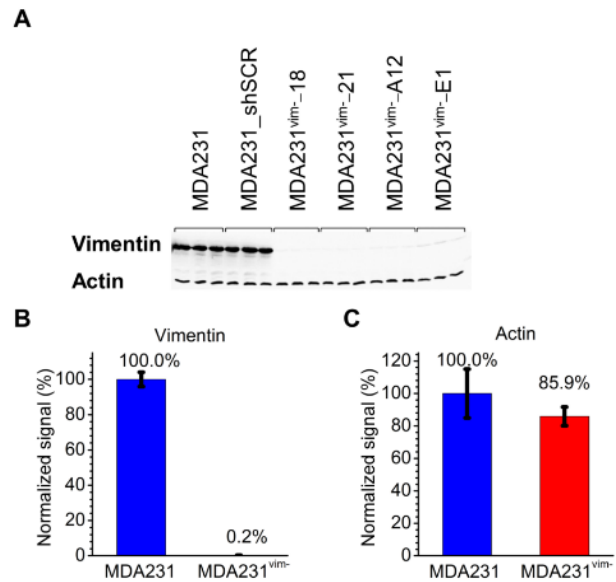

**Figure S1. Western blots.** (A) Western blot gel bands for MDA231 and the various MDA231$^{vim-}$ clones. (B-C) Quantitative analysis of the vimentin (B) and actin bands (C), normalized by the median read of the MDA231 cells. Error bars represent the standard error of the mean (SEM). n=6 for MDA231 and n=18 for MDA231$^{vim-}$.

## II. Exponential velocity distribution and cell heterogeneity

Velocity distribution was assessed from the cell tracking data. The complementary cumulative distribution function $F(v) = \int_0^v P(v')dv'$, $P(v)$ being the velocity probability distribution function, is displayed in Fig. S2A-B for various experiments. $1 - F(v)$ shows a good fit to an exponential distribution (equivalent to $P(v)$ being exponentially distributed, since in that case: $F(v) = \int_0^v \frac{1}{\bar{v}} e^{-v'/\bar{v}} dv' = 1 - e^{-v/\bar{v}}$) in a robust manner for both the MDA231 and MDA231$^{\text{vim-}}$ cell lines, as well as for the whole density range tested in our experiments (70-600 cells/mm$^2$). Furthermore, velocity distributions of single cells exhibited the same functional dependence, indicating that the exponential distribution is inherent for the studied cells, rather than reflecting heterogeneity in cell populations (Fig. S2C). These findings are in agreement with previous results for various cell lines, including wild-type MDA231 [5]. However, we determined that during their lifetimes, cells fluctuate around an average velocity unique to each cell, rather than around the ensemble average. Fig. S2D shows the autocorrelation of the deviation of cell speed from the ensemble average: $C_{\Delta v}(t) = \langle \Delta v_i(\tau) \Delta v_i(\tau + t) \rangle$, where $\Delta v_i(\tau) = v_i(\tau) - \bar{v}(\tau)$, and $<>$ is the ensemble and time average. $C_{\Delta v}(t)$ remains positive over the entire lifetime of a cell (approximately 1 day=1440 min). This means that in addition to the environmentally mediated effects on cell speed discussed in this work, there might be further long-term intrinsic processes that determine cell speed.

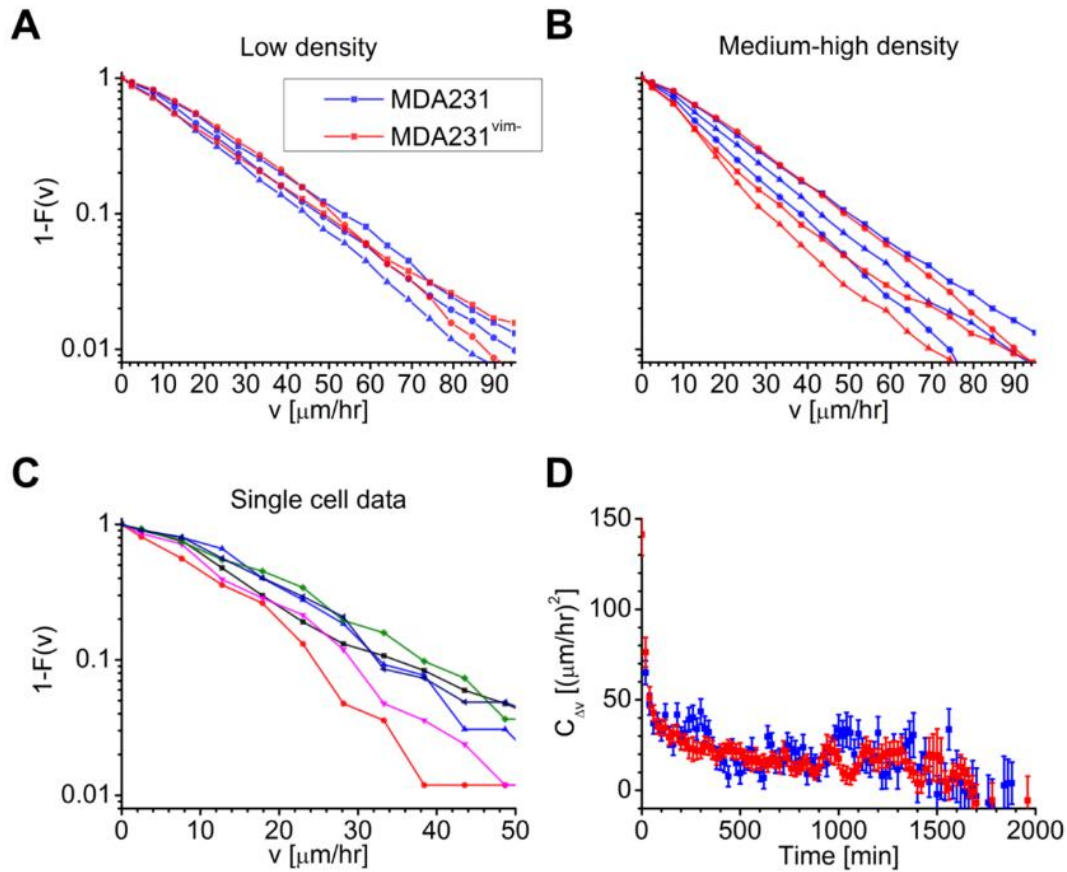

**Figure S2. Cell Velocity Distribution**. (A-B) Velocity distribution functions of single experiments, at low (<120 cells/mm$^2$) (A) and medium-high (200-600 cells/mm$^2$) densities (B). Both MDA231 and MDA231$^{vim-}$ cell lines display exponentially distributed velocities, regardless of plate density. (C) Velocity distribution functions of 5 individual cells in a low density MDA231 experiment. The velocities of single cells also distribute exponentially, even though data is noisier, due to a smaller sample size. (D) Time autocorrelation of velocity fluctuations of single cells from the ensemble average. The autocorrelation does not decay to zero during the scale of a cell's lifetime (1440 min), indicating that being fast or slow is an intrinsic cell property.

### III. Cell proliferation rate

Differences in division rate between the MDA231 and MDA231$^{vim-}$ cell lines could impact assays such as the wound-healing assay, where a faster division rate could enable cells to reach confluency in the wound. Therefore, XTT–based proliferation assays were performed. Our results rule out such differences, indicating that no significant difference exists between the two cell lines (Fig. S3).

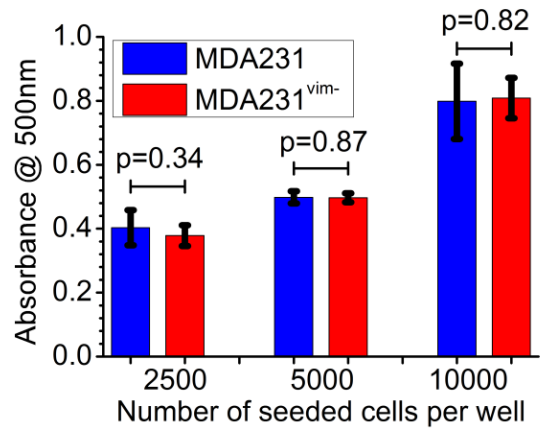

**Figure S3. Cell proliferation analysis.** MDA231 and MDA231$^{vim-}$ cells were seeded at 3 different concentrations, and allowed to proliferate for 24 hrs. An assay reaction solution was then added for 3 hours, and absorbance intensity was measured. No difference in proliferation rate between MDA231 and MDA231$^{vim-}$ was measured. Error bars represent the SEM.

## IV. Local density calculation

In order to capture the density heterogeneity within a plate, we calculated the local density at every point. The local density can be thought of as the sum of additive fields $\phi(r)$ generated by each cell, $r$ being the distance of each cell from the point. $\phi(r)$ should decay to zero at a large distance from the cell; the relevant calculation (see also Fig. S4) is described by the equation:

$$\phi(r) = \begin{cases} 1 & r \leq d_a \\ 2\dfrac{d_a}{r} - 1 & d_a < r < 2d_a \\ 0 & r \geq 2d_a \end{cases}$$

Here, we use $d_a = 50\mu m$, which is approximately one cell diameter. Thus, the local density at point $\vec{r}$ can be calculated as:

$$\rho_{local}(\vec{r}) = \sum_i \phi(|\vec{r} - \vec{r}_i|)$$

where the summation is over the entire population of cells within the frame, $\vec{r}_i$ being the coordinate of each cell.

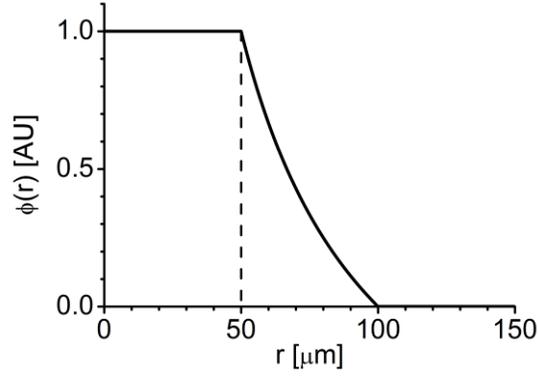

**Figure S4. Local density field.** Local density field, for a cell located at a distance r from the considered point. The local density field saturates at $r = 50\mu m$ (approximately one cell diameter), and decays to zero at $r = 100\mu m$. The local density at a point $\rho_{local}$ is the sum of the fields generated by all cells within the frame.

### V. Arrangement of cells at high density

The arrangement of cells in a plate can be evaluated by the pair-correlation function, defined as:

$$g(r) = \left\langle \frac{V}{N-1} \sum_i \frac{\delta(r - r_i)}{2\pi r dr} \right\rangle$$

Here, for each cell we sum over the rest of the cells distanced at $r_i$, $\delta$ is Kronecker's delta function, and $r$ is the radial coordinate, with the chosen cell set as the origin. Thus, the number of cells in the annulus $r \leq R \leq r + \delta r$ around the center cell is obtained, and it is normalized by the annulus' area ($2\pi r dr$). A further time and ensemble average, denoted by $<>$, is performed over N center cells for the total V imaged area. As shown in Fig. S5, $g(r)$ peaks at a typical length scale for both cell lines, with the length being shorter for MDA231[vim-] cells (47 μm for MDA231, vs.

42 μm for MDA231$^{vim-}$). This result suggests that MDA231$^{vim-}$ cells prefer to adhere to each other more tightly. Alternatively, it could reflect the fact that MDA231 cells are more motile, as after cell division, daughter cells can separate more rapidly from each other. This could lead the MDA231 cells to form a more homogenous monolayer over time. On the other hand, a smaller typical distance could represent a stand-alone phenomenon, due to tighter cell-cell adhesions.

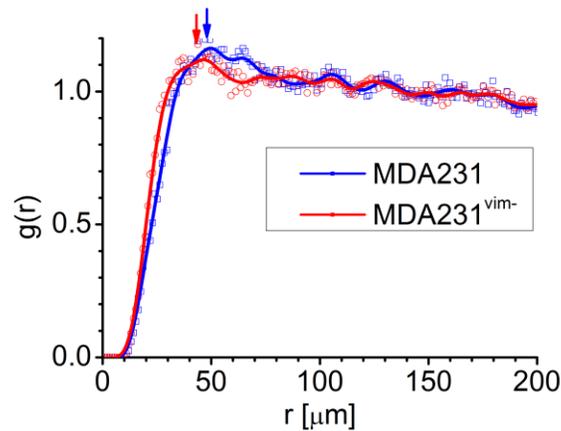

**Figure S5. Pair correlation function at dense plates**. Averaged g(r) (over 4 curves for both MDA231 and MDA231$^{vim-}$) for experiments at high density (>400 $\frac{cells}{mm^2}$). Data is marked by circles, lines are calculated by FFT smoothing on a 5.48 μm window. For both cell lines, $g(r)$ remains 0, up to about 10 mm. Arrows mark peaks at 47 μm and 43 μm for MDA231 and MDA231$^{vim-}$, respectively, indicating the preferred cell-to-cell distance.

## VI. Infiltration assay leading edge velocities

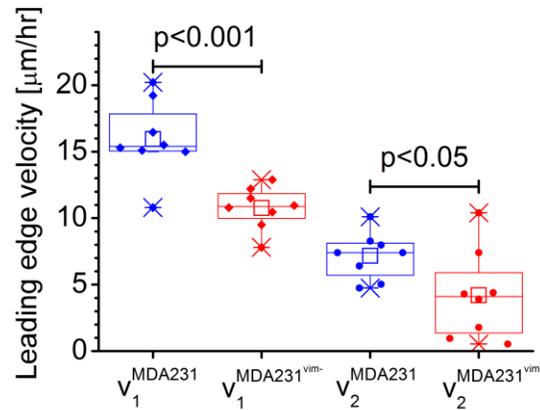

**Figure S6. Infiltration assay leading edge velocities.** Leading edge velocities of MDA231/MDA231[vim-] monolayers at infiltration assays (see Fig. 5 caption for a description of the calculation procedure). $v_1, v_2$ correspond to the migration before and after contact with the stroma, respectively. Mean ± SEM (n=8 for both cell lines): $v_1^{MDA231} = 16.0 \pm 1.0 \frac{\mu m}{hr}$, $v_1^{MDA231^{vim-}} = 10.8 \pm 0.6 \frac{\mu m}{hr}$, $v_2^{MDA231} = 7.2 \pm 0.6 \frac{\mu m}{hr}$, $v_2^{MDA231^{vim-}} = 4.2 \pm 1.2 \frac{\mu m}{hr}$.

## VII. Coherent cell motility

Coherent cell migration can enhance cell motility in dense cell cultures [6]. Merely reorienting toward a coherent direction of migration prevents neighboring cells from obstructing each other's movements. Therefore, one hypothesis we considered is that MDA231[vim-] cells lose this capacity to some extent, impairing their motility at high density. However, searching for signatures of coherent migration by means of flow field analysis described by Szabó et al. [7] (see Fig. S7A for a schematic explanation) shows that coherent migration is very short-ranged (in terms of distance), and similarly weak in MDA231 and MDA231[vim-] cell lines (Fig. S7B; Fig. S8). These results imply that cells of both cell lines migrate individually, without correlation with their neighbors' movements, and that inter-cellular cooperation does not constitute a distinctive factor between these cell lines.

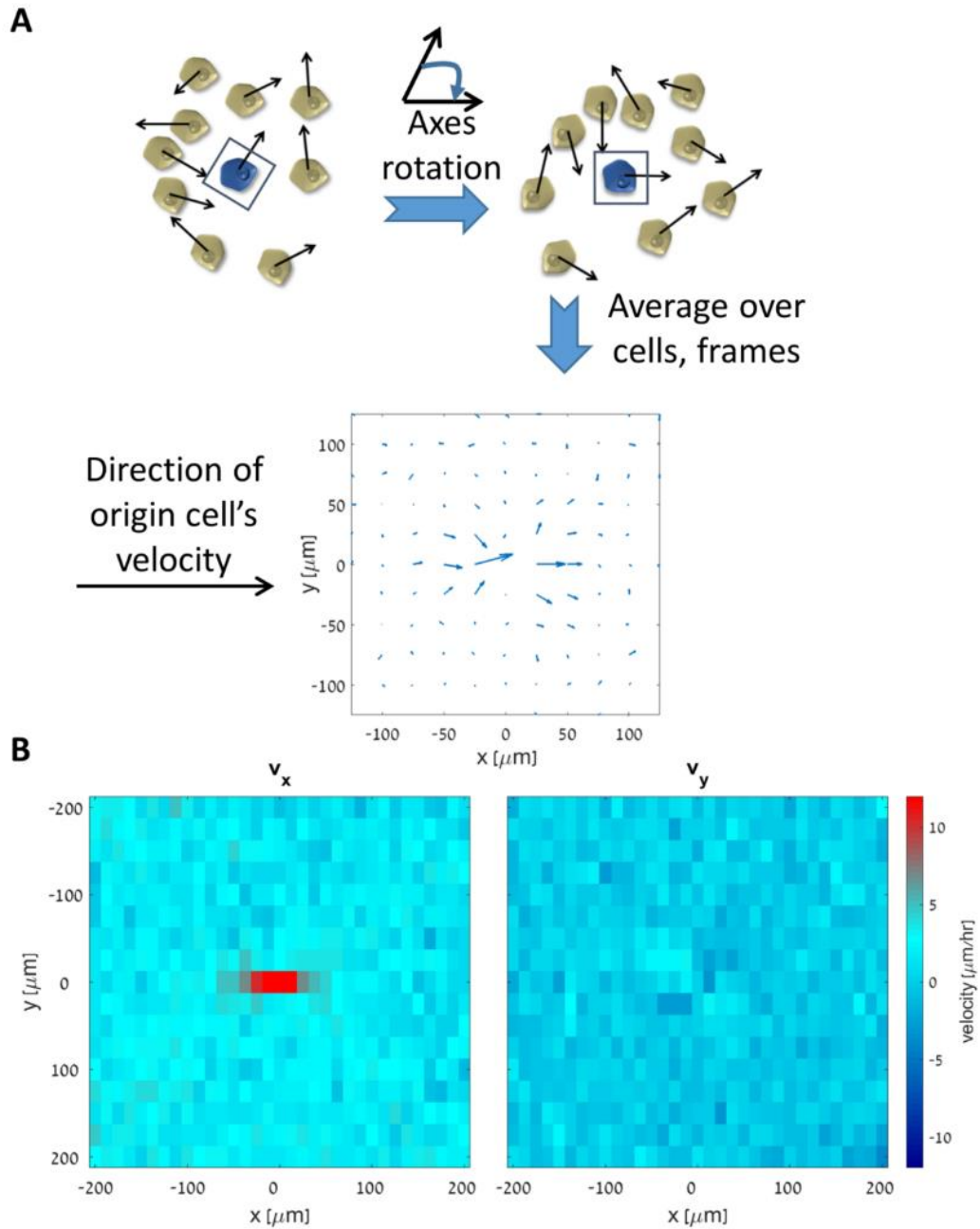

**Figure S7. Flow field calculation.** (A) Each cell is chosen as the origin of the axes. The frame of reference is rotated so that the origin cell's velocity lies in the positive x direction. The neighboring
42

cells' velocities are associated with their relative displacements from the origin cell. Averaging over cells and frames gives the flow field, which is a vector field of velocities likely to be seen when moving with a cell's frame of reference. The results show short-range coherent movement in the axis of movement. (B) Maps of the x,y (parallel and perpendicular) components of the flow field. For $v_x$, the short-range correlation is noticeable at a narrow strip up to ±50 μm from the cell, in the axis of movement. For $v_y$, no correlation is found.

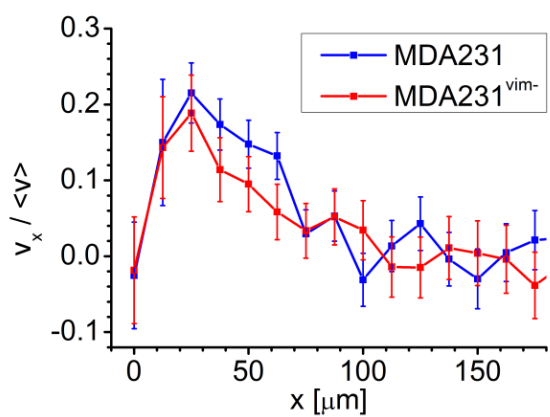

**Figure S8. Flow field analysis results**. $v_x$ is the projection of a neighboring cell's velocity on the direction of the origin cell's velocity. $v_x$ is associated with the $x$ coordinate, a projection of the displacement between the two cells in the origin cell's heading direction. $v_x$ is normalized by the median cell velocity in the plate $\langle v \rangle$. The ensemble and time average produce the flow field profile of an experiment. Curves show averaged data for 3 MDA231 and 2 MDA231$^{vim-}$ high density experiments; data was taken from the narrow strip $y = [-31.25\mu m, 31.25\mu m]$. Both cell lines display a weak coherent movement component, decaying in the short range of approximately one cell diameter (50 μm), with a slightly more significant component for the MDA231 cells.

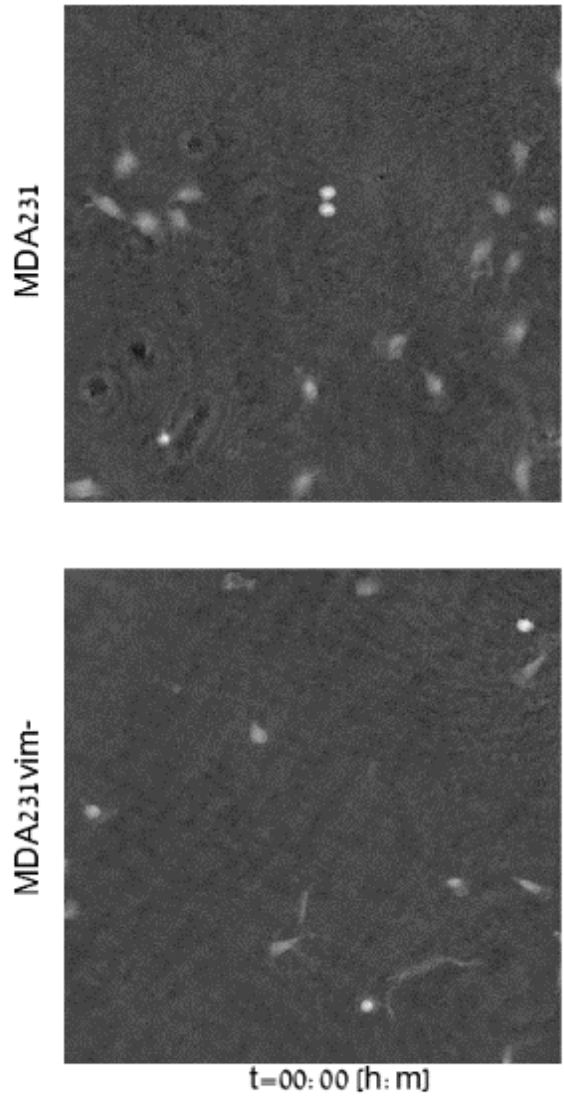

**Movie 1. Cell tracking with the Holomonitor M4 device.** The fields shown are at low cell density, and correspond to 400 min.

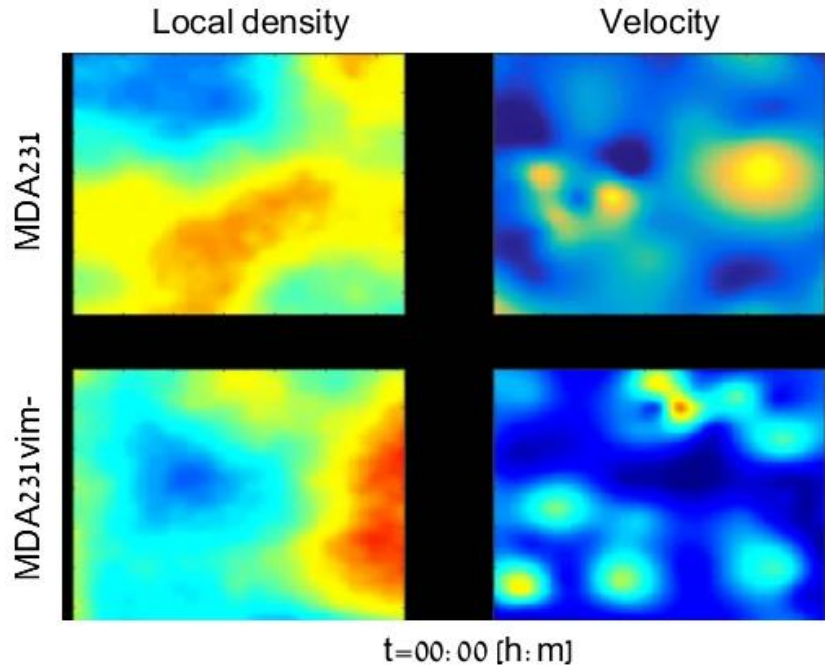

**Movie 2**. **Velocity and local density evolution over time for MDA231 and MDA231$^{vim-}$ cells.** Local density (left) is calculated as detailed in Supporting Information, Section IV. Velocity field (right) is calculated using MATLAB® biharmonic spline interpolation for the momentary velocity of each cell. Hot/cool colors indicate high/low values (color map is the same as in Fig. 3C).

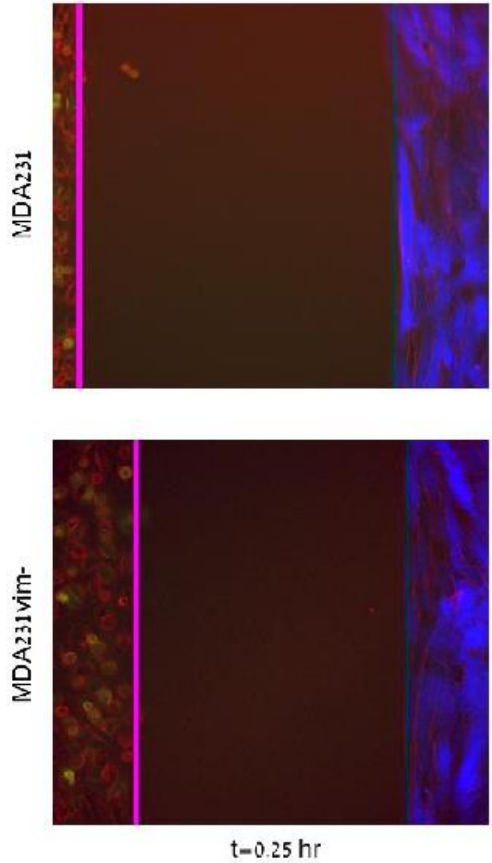

**Movie 3. Infiltration to stroma assay.** MDA231/MDA231$^{vim-}$ cells (colored in green on the left) are seeded opposite to fibroblast stroma cells (colored in blue on the right). Magenta and cyan lines show the edge layers, as detected automatically by an algorithm.

**Supplementary References**